\documentclass[10pt,journal]{IEEEtran}

\usepackage{graphicx}
\usepackage{amsmath}
\usepackage{amssymb}
\usepackage{xcolor}
\newcommand{\divkpi}{model diversity}   
\newcommand{\rv}[1]{\textcolor{black}{#1}}
\usepackage{bm}
\usepackage{multirow}
\usepackage{booktabs}
\usepackage{algorithm}
\usepackage{algpseudocode}
\usepackage{pifont}
\usepackage{xcolor}
\usepackage{hyperref}
\usepackage{placeins}
\usepackage{cite}
\usepackage{authblk}

\newcommand{\etal}{\textit{et al.}}
\begin{document}

\title{Fuzzy-FedAvg: Server-Aware Client Selection for Federated Speech Emotion Recognition in Heterogeneous Environments}

\author{
Farrukh Pervez$^{1}$, Siddique Latif$^{2}$, Moid Sandhu$^{\dag 3,2}$\thanks{$^\dag$Corresponding Author: Moid Sandhu (e-mail: moid.sandhu@csiro.au)}, Raja Jurdak$^{2}$
}

\affil{${^1}$College of Aeronautical Engineering, National University of Sciences \& Technology (NUST), Islamabad 24090, Pakistan}
\affil{$^{2}$Queensland University of Technology (QUT), Brisbane QLD 4000, Australia}
\affil{$^{3}$Commonwealth Scientific and Industrial Research Organisation (CSIRO), Brisbane, QLD 4029, Australia}

\maketitle

\begin{abstract}

\color{black}{Federated speech emotion recognition (SER) is challenged by speaker-dependent, class-imbalanced data and by heterogeneous client devices. This paper proposes \textit{Fuzzy-FedAvg}, a server-aware framework in which the server learns Client Selection Weights (CSWs) through fuzzy Q-learning from global-model feedback. The same CSWs weight client accuracy, training time, and \divkpi{} in both client selection and aggregation, while a dedicated PCAS node manages stragglers through staleness-discounted deferred updates. On IEMOCAP and CREMA-D, \textit{Fuzzy-FedAvg} achieves higher unweighted average recall (UAR) than FedAvg in all configurations under severe non-IID conditions, and in 22 of 24 participation-level and 10 of 12 training-horizon configurations overall. It also outperforms FedProx and Oort, while SCAFFOLD and Power-of-Choice remain competitive, and it reduces cumulative training time by 81.2\% in a FedScale-based evaluation. The benefit is greatest under strong statistical heterogeneity and narrows as data become more balanced.}
\end{abstract}


\section{Introduction}
Speech Emotion Recognition (SER) aims to detect and interpret human emotions from speech and has gained significant traction across a broad range of real-world domains, including healthcare diagnostics, customer service, smart virtual assistants, education, and entertainment \cite{khan2026generative, chhikara2020federated,girardi2026role, latif2026affective,pervez2024medicine,latif2024evaluating,sandhu2026empowering}. {\color{black}Applications range from passive monitoring of mental-health conditions such as depression and anxiety, to real-time sentiment detection in customer service, adaptive learning platforms that respond to student frustration, and emotionally aware virtual assistants.} Despite this broad utility, traditional SER systems depend on collecting and centralising large volumes of users' raw speech data for model training, which raises serious and multifaceted privacy concerns \cite{dar2024speech,latif2025transformers,latif2025can,ali2022multi}. Speech signals inherently carry a dense spectrum of sensitive information --- including age, gender, biometric identity, personality traits, location, emotional state, and potential health indicators --- making centralised data aggregation not only vulnerable to direct misuse but also susceptible to unintended inference attacks whereby adversaries can extract attributes far beyond what users consented to share \cite{kroger2019privacy,zaidi2024enhancing,latif2020deep}. {\color{black} This combination of data sensitivity and inference risk makes the centralised training paradigm difficult to reconcile with the privacy expectations of modern users and the obligations imposed by data protection frameworks \cite{pervez2024affective,ali2021privacy}.}

Federated Learning (FL) has emerged as a compelling paradigm to address these privacy concerns by enabling geographically distributed users to collaboratively train a shared global model without transferring their raw speech data to a central server \cite{mcmahan2017communication, konevcny2016federated, tripathi2019deep}. {\color{black} Rather than pooling sensitive data in a single location, FL confines raw information to the local devices of each participating client, with only model parameters or gradient updates transmitted to the server, thereby reducing the attack surface associated with centralised raw-data storage \cite{sandhu2024exploring}.} {\color{black} This design aligns with} core data-protection principles such as data minimisation and purpose limitation, and supports institutional compliance with regulations such as the General Data Protection Regulation (GDPR) \cite{protection2018general}. {\color{black}In each global round, participating clients train the current global model on their private data and return their updates, which the server aggregates, typically through weighted averaging, into a new global model; this process repeats until convergence.}

Federated Averaging (FedAvg) is {\color{black} the most widely used} framework in FL environments \cite{dhawan2023leveraging}. It is communication-efficient, as model weights are transmitted to the central server only after {\color{black} multiple epochs of local training} on edge devices. {\color{black} While FedAvg retains raw training data on local devices and reduces communication overhead, the exchange of model updates may still expose client information. Furthermore, its convergence can become slow and unstable under non-independent and identically distributed (non-IID) speech data distributions \cite{li2019convergence, cao2025comparative,shoukat2023breaking,latif2023generative,malik2023preliminary}.} In addition, FL systems are affected by the straggler problem, where slow or resource-constrained clients delay the global aggregation process, thereby increasing overall training time \cite{li2019convergence,latif2023survey}.

\begin{table*}[!ht]
\centering
\caption{\rv{Comparison of \textit{Fuzzy-FedAvg} with related federated SER, general FL, and decentralised FL studies.}}
\label{tab:literature_comparison}
\scriptsize
\setlength{\tabcolsep}{4pt}
\renewcommand{\arraystretch}{1.1}
\begin{tabular}{p{2.2cm} p{1cm} p{2.5cm} p{2.5cm} p{2.5cm} p{2.2cm} p{2cm}}
\hline
\textbf{Paper} & \textbf{Year} & \textbf{Dataset(s)} & \textbf{Feature Space} & \textbf{Client Selection Scheme} & \textbf{Federated Environment} & \textbf{Heterogeneity Level} \\
\hline
\multicolumn{7}{l}{\rv{\textit{Federated SER}}} \\
\hline
Latif et al. \cite{latif2020federated} & 2020 & IEMOCAP & Mel Filterbank & None & Homogeneous & N/A \\
Somandepalli et al. \cite{somandepalli2022federated} & 2022 & IEMOCAP & Utterance-level acoustic features & None & Homogeneous & N/A \\
Tsouvalas et al. \cite{tsouvalas2022privacy} & 2022 & IEMOCAP & Mel & None & Heterogeneous & Statistical heterogeneity (speaker variability) \\
Feng \rv{et al.} \cite{feng2022semi} & 2022 & IEMOCAP, \rv{MSP-Improv} & EmoBase feature set, APC, Tera, DeCoAR 2.0, DistilHuBERT \& MFCC & None & Heterogeneous & Statistical heterogeneity (non-IID) \\
Feng \rv{et al.} \cite{feng2022user} & 2022 & IEMOCAP, CREMA-D, MSP-Improv & EmoBase and APC feature set & None & Homogeneous & N/A \\
\rv{Mohammadi} et al. \cite{mohammadi2023balancing} & 2023 & CREMA-D & EmoBase feature set & Half of the clients are selected based on dataset size and remaining half randomly & Homogeneous & N/A \\
Zhang et al. \cite{zhang2023fedaudio} & 2023 & IEMOCAP, CREMA-D & APC feature set, Mel & None & Homogeneous & N/A \\
Chowdhury et al. \cite{chowdhury2023privacy} & 2023 & IEMOCAP, CREMA-D & MFCC & None & Homogeneous & N/A \\
Mishchenko et al. \cite{mishchenko2023hyperparameters} & 2023 & CREMA-D & Not mentioned & None & Heterogeneous & Statistical heterogeneity (non-IID) \\
Feng et al. \cite{feng2024partial} & 2024 & IEMOCAP, CREMA-D & Mel Filterbank & None & Homogeneous & N/A \\
\hline
\multicolumn{7}{l}{\rv{\textit{General FL}}} \\
\hline
\rv{FedProx \cite{li2020fedprox}} & \rv{2020} & \rv{Non-SER benchmarks} & \rv{N/A} & \rv{None (random)} & \rv{Heterogeneous} & \rv{Statistical and system heterogeneity} \\
\rv{SCAFFOLD \cite{karimireddy2020scaffold}} & \rv{2020} & \rv{Non-SER benchmarks} & \rv{N/A} & \rv{None (random)} & \rv{Heterogeneous} & \rv{Statistical heterogeneity (client drift)} \\
\rv{FedCS \cite{nishio2019client}} & \rv{2019} & \rv{Non-SER benchmarks} & \rv{N/A} & \rv{Deadline-based, resource-aware selection} & \rv{Heterogeneous} & \rv{System heterogeneity} \\
\rv{Oort \cite{lai2021oort}} & \rv{2021} & \rv{Non-SER benchmarks} & \rv{N/A} & \rv{Utility-based selection combining statistical and system utility} & \rv{Heterogeneous} & \rv{Statistical and system heterogeneity} \\
\rv{Power-of-Choice \cite{cho2022client}} & \rv{2022} & \rv{Non-SER benchmarks} & \rv{N/A} & \rv{Selection biased towards higher local loss} & \rv{Heterogeneous} & \rv{Statistical heterogeneity (non-IID)} \\
\rv{FAVOR \cite{wang2020optimizing}} & \rv{2020} & \rv{Non-SER benchmarks} & \rv{N/A} & \rv{Deep Q-learning-based selection} & \rv{Heterogeneous} & \rv{Statistical heterogeneity (non-IID)} \\
\rv{FedAsync \cite{xie2019asynchronous}, FedBuff \cite{nguyen2022federated}} & \rv{2019, 2022} & \rv{Non-SER benchmarks} & \rv{N/A} & \rv{Asynchronous participation with staleness-weighted updates} & \rv{Heterogeneous} & \rv{System heterogeneity} \\
\hline
\multicolumn{7}{l}{\rv{\textit{Decentralised FL}}} \\
\hline
\rv{DT-DOFL \cite{jiang2025dt}, DeFedHDP \cite{wei2024defedhdp}, AccDFL \cite{wei2024accdfl}, DePsFL \cite{zhao2025differential}} & \rv{2024--2025} & \rv{Healthcare and non-SER data} & \rv{N/A} & \rv{None (peer-to-peer coordination)} & \rv{Decentralised} & \rv{Not the primary focus} \\
\hline
\rv{\textbf{\textit{Fuzzy-FedAvg} (Proposed)}} & 2026 & IEMOCAP, CREMA-D & MFCC, Mel\rv{-spectrogram} & Server-driven selection using dynamically learned KPI-based CSWs & Heterogeneous & Statistical (non-IID) and system heterogeneity \\
\hline
\end{tabular}
\end{table*}

{\color{black} Federated SER presents challenges beyond conventional non-IID learning because emotional speech exhibits substantial speaker-dependent variability in vocal expression, speaking style, and acoustic characteristics. Moreover, emotion classes are often unevenly represented across speakers, sessions, and recording conditions, resulting in heterogeneous local data distributions. Consequently, clients may contribute substantially different information to the global SER model, even when they hold comparable amounts of training data. Conventional FedAvg relies on random client selection without explicitly accounting for differences in local model quality, \divkpi{}, or execution time. Several adaptive client-selection strategies have been proposed for general FL, including resource-aware selection \cite{nishio2019client}, utility-guided participant selection \cite{lai2021oort}, and reinforcement-learning-based selection \cite{wang2020optimizing}. However, these approaches typically rely on fixed selection criteria, treat client selection and model aggregation as independent steps, or discard updates from delayed clients, and, to the best of our knowledge, none has been examined for the speaker-dependent heterogeneity of federated SER. These characteristics motivate an adaptive client-selection mechanism that considers the contribution and availability of individual clients throughout federated SER training.}

Accordingly, we propose {\color{black} \textit{Fuzzy-FedAvg}, a server-aware, time-efficient client-selection scheme for federated SER under non-IID data.} Clients report {\color{black} Key Performance Indicators (KPIs)} --- including accuracy, training time, and {\color{black}\divkpi{}} --- to the server via a PCAS (Profiling, Client Selection and Aggregation) node{\color{black}, which also performs model aggregation and handles communication between the server and the selected clients}. The server is equipped with a fuzzy Q-learning mechanism that dynamically adapts its Client Selection Weights (CSWs) based on the global model KPIs observed in each training round{\color{black}, and communicates them to the PCAS node to guide subsequent selection}. {\color{black} Fuzzy Q-learning is well suited to this setting: the fuzzy inference layer maps noisy, continuous global-model KPIs into interpretable linguistic states, while Q-learning adapts the CSWs through trial-and-error interaction without requiring a prior model of client behaviour or data distribution. Hence, the term \textit{server-aware} refers to the adaptation of client-selection priorities by the server based on global-model feedback, while client-level KPIs are used by the PCAS node to rank and select clients according to these priorities.}

{\color{black} Unlike approaches that employ fixed selection criteria or configure selection and aggregation independently, \textit{Fuzzy-FedAvg} uses a single set of learned CSWs to govern both operations, complemented by a three-tier straggler-management mechanism that retains eligible delayed updates with an explicit staleness discount. A comparison with existing studies is presented in Table~\ref{tab:literature_comparison}, and the main contributions of this paper are summarised as follows:
\begin{itemize}
\item We propose \textit{Fuzzy-FedAvg}, a server-aware framework for federated SER that employs a dedicated PCAS node to decouple client profiling, selection, and aggregation from the server's learning role, allowing the server to focus solely on adapting selection priorities from global-model feedback.
\item We develop an adaptive learning and aggregation mechanism in which fuzzy Q-learning adjusts the CSWs based on global-model feedback. The dynamically learned CSWs govern both client ranking and model aggregation, establishing a coordinated relationship between participant selection and update contributions.
\item We incorporate a three-tier, threshold-based straggler-management mechanism that distinguishes timely, deferred, and straggling clients. Eligible delayed updates are retained for the subsequent round and incorporated through staleness-discounted aggregation, reducing the loss of completed local computations.
\item We extend client-selection research to federated SER under simultaneous statistical and system heterogeneity, moving beyond the dataset-size-based and random selection previously investigated for SER \cite{mohammadi2023balancing}. Experiments on IEMOCAP and CREMA-D show that \textit{Fuzzy-FedAvg} improves UAR over FedAvg, FedProx, and Oort under severe non-IID conditions and reduces cumulative training time by 81.2\% in a FedScale-based evaluation.
\end{itemize}
}

{\color{black} The remainder of this paper is organised as follows. Section~\ref{sec:related_work} reviews related work on federated SER and client selection. Section~\ref{sec:methodology} presents the system model, the problem formulation, and the proposed \textit{Fuzzy-FedAvg} framework. Section~\ref{sec:experimental_setup} describes the experimental setup, and Section~\ref{sec:results} reports the results. Finally, Section~\ref{sec:conclusion} discusses the findings and concludes the paper.}

\section{Related Work}
\label{sec:related_work}
{\color{black}Beyond privacy, a central difficulty in federated SER is} that emotional speech data is inherently speaker-dependent, making non-IID distributions the rule rather than the exception, and that the devices holding this data differ radically in computational capacity and network reliability. Standard federated aggregation degrades under either condition; under both simultaneously, it has not been adequately studied in the SER literature. Latif~\etal~\cite{latif2020federated} and Somandepalli~\etal~\cite{somandepalli2022federated} established the earliest federated baselines for SER, demonstrating that FedAvg achieves parity with centralised training on IEMOCAP under IID conditions. The limitation is structural: IID assumes each client holds a representative sample of the global distribution, {\color{black}a condition rarely met in realistic multi-speaker deployments}. Both works acknowledge this boundary; neither crosses it.

Subsequent efforts confronted the label scarcity and distribution skew that characterise real on-device emotional data. Tsouvalas~\etal~\cite{tsouvalas2022privacy} embedded self-training within the FL loop to exploit unlabelled local samples, recovering 8\% in average unweighted accuracy over fully supervised federated models under speaker-variable non-IID conditions --- a gain attributable to reducing the effective data imbalance across clients rather than to any change in the aggregation scheme itself. Feng~\etal~\cite{feng2022semi} pushed further with semi-FedSER, showing that performance remains competitive when labelled data accounts for only 20\% of local samples across IEMOCAP and MSP-Improv. Taken together, these results make explicit what FedAvg-based baselines obscured: statistical heterogeneity is a first-order performance variable in federated SER, and reducing it --- rather than accommodating it --- has been the operative strategy. Neither work touches the question of which clients should participate or when.

Privacy introduces a second constraint that interacts non-trivially with heterogeneity. Feng~\etal~\cite{feng2022user} showed that user-level differential privacy suppresses attribute inference attacks effectively, but the accuracy cost is substantial{\color{black}, and this cost can be expected to grow under non-IID conditions, where noise introduced by privacy mechanisms compounds the existing divergence between local models}. Mohammadi~\etal~\cite{mohammadi2023balancing} attempted to recover some of this loss by coupling local differential privacy with a client selection rule: half the participants chosen by dataset size, the remainder at random. This is the only prior SER work that explicitly controls who trains in each round, and it does reduce selection bias. But the rule is fixed --- it does not observe model behaviour, respond to per-round data quality, or account for device capability --- so it cannot adapt to the conditions it was designed to handle.

Three other directions have refined the federated SER picture without addressing heterogeneity directly. Zhang~\etal~\cite{zhang2023fedaudio} showed through the FedAudio benchmark that label noise and acoustic corruption inflict disproportionate damage on federated models relative to centralised ones, because corrupt local updates propagate into the global model without the error-averaging that large centralised datasets provide; this finding establishes data quality as a dimension any serious client selection policy must account for. Chowdhury~\etal~\cite{chowdhury2023privacy} found that CNN and LSTM models under standard FedAvg outperform attention-based federated schemes on IEMOCAP and CREMA-D{\color{black}, suggesting that aggregation quality, rather than architectural sophistication, is a key constraint on performance}. Mishchenko~\etal~\cite{mishchenko2023hyperparameters} improved both accuracy and wall-clock training time through joint server--client hyperparameter optimisation under non-IID conditions, but left the participation set unchanged across rounds. Feng~\etal~\cite{feng2024partial} introduced partialFL, allowing selective data sharing with the server and reporting competitive results against FedAvg and FedProx on multi-modal data; the evaluation, however, assumes homogeneous system environments throughout, leaving device-level variability as an open variable.

{\color{black}
Outside SER, statistical and system heterogeneity have been studied extensively in general FL. FedProx~\cite{li2020federated} adds a proximal term to limit local model drift under non-IID data, and SCAFFOLD~\cite{karimireddy2020scaffold} uses control variates to correct client drift; both modify local optimisation but leave client participation random. Client-selection methods instead target which clients train in each round. FedCS~\cite{nishio2019client} selects clients that can complete training within a deadline, Oort~\cite{lai2021oort} ranks clients by combining statistical utility with system speed, and Power-of-Choice~\cite{cho2022client} biases selection towards clients with higher local loss to accelerate convergence. FAVOR~\cite{wang2020optimizing} learns a selection policy through deep Q-learning to counteract non-IID bias. For stragglers, asynchronous schemes such as FedAsync~\cite{xie2019asynchronous} and FedBuff~\cite{nguyen2022federated} incorporate delayed updates using staleness-dependent weighting.

Decentralised federated learning (DFL) takes a different direction by removing the central server, with clients exchanging updates directly among peers. DT-DOFL~\cite{jiang2025dt} introduces a digital-twin-assisted framework for decentralised online federated learning in healthcare, incorporating DP-DOOR as a differentially private online learning algorithm. DeFedHDP~\cite{wei2024defedhdp} combines decentralised online aggregation with differential privacy for heart-disease prediction, while AccDFL~\cite{wei2024accdfl} investigates accelerated decentralised learning in healthcare IoT networks. DePsFL~\cite{zhao2025differential} further combines differential privacy, push-sum coordination, and two-point gradient estimation over time-varying networks. These approaches primarily address decentralised model coordination, optimisation efficiency, and privacy.
}

{\color{black}
\subsection{Summary and Contribution}
\label{subsec:summary_contribution}
As summarised in Table~\ref{tab:literature_comparison}, three gaps emerge from this literature. First, among the federated SER studies, seven of ten assume homogeneous federated environments, the remaining three address statistical heterogeneity only, and none considers system heterogeneity. Client participation has either been left unmanaged or governed by a fixed rule; the strategy in~\cite{mohammadi2023balancing}, for instance, combines dataset-size-based and random sampling without incorporating client performance, \divkpi{}, or execution time. Second, general FL selection methods adapt participation, but they typically optimise a single criterion or a fixed combination of criteria, treat selection and aggregation as separate steps, and have not been evaluated on the speaker-dependent heterogeneity of emotional speech. Third, decentralised FL approaches investigate distributed coordination, optimisation, and privacy rather than adaptive participant selection, while asynchronous methods retain delayed updates without adapting which clients are selected.

\textit{Fuzzy-FedAvg} addresses these gaps through a server-coordinated framework that jointly considers statistical and system heterogeneity in federated SER. Fuzzy Q-learning learns KPI-based CSWs from global-model feedback, replacing fixed selection rules with priorities that adapt across training rounds. The same CSWs govern both client ranking and model aggregation, linking which clients participate with how strongly each update contributes to the global model. A dedicated PCAS node implements these decisions and manages delayed updates through threshold-based three-tier classification and staleness-discounted aggregation, so that completed local computation is not discarded. As the final row of Table~\ref{tab:literature_comparison} shows, \textit{Fuzzy-FedAvg} is the only SER approach that combines adaptive client selection with both statistical and system heterogeneity, and it is evaluated on IEMOCAP and CREMA-D using MFCC and Mel-spectrogram features.
}

\section{Methodology}
\label{sec:methodology}

{\color{black}
This section first describes the federated SER system and
formulates the client-selection problem, and then presents the
proposed \textit{Fuzzy-FedAvg} solution.
}

\subsection{System Model and Problem Formulation}
\label{sec:system_model}
\label{subsec:arch}

\subsubsection{System Model}

We consider a heterogeneous FL system comprising three logical
entities: a central server~$S$, a \emph{PCAS} node, and $K$~distributed clients,
each holding a private local dataset
$\mathcal{D} = \{\mathcal{D}_1, \mathcal{D}_2, \dots,
\mathcal{D}_K\}$. The clients collaboratively train a shared
global SER model without ever transferring raw speech data to
either the server or the PCAS node, thereby preserving data
locality and adhering to the data-minimisation principle~\cite{mcmahan2017communication}.
{\color{black}In each global round, a subset of $N \leq K$ clients
trains on the current global model and returns its updates for
aggregation; the roles of the server and the PCAS node in this
process are detailed in Section~\ref{subsec:solution}.}

{\color{black}
Real-world federated SER deployments exhibit two compounding
forms of heterogeneity. The first is \emph{statistical
heterogeneity}: clients hold speech recorded under different
acoustic conditions and from speakers with distinct demographic
profiles, giving rise to non-IID local data distributions. We
model this by partitioning the global dataset among the $K$
clients using a Dirichlet distribution~\cite{yurochkin2019bayesian},
the standard mechanism for generating non-IID partitions in FL
benchmarks~\cite{li2022federated}:
\begin{equation}
f(\eta_1, \dots, \eta_K;\, \alpha)
= \frac{\Gamma(\alpha \cdot K)}{\Gamma(\alpha)^K}
  \prod_{i=1}^{K} \eta_i^{\,\alpha - 1},
\label{eq:dirichlet}
\end{equation}
where $\eta_i$ is the probability of assigning an emotion sample
to client~$i$, $\Gamma(\cdot)$ is the gamma function, and
$\alpha$ is the concentration parameter controlling label skew.
A small value such as $\alpha = 0.1$ produces highly skewed
partitions in which each client's corpus is dominated by one or
two emotion classes, as in a contact-centre or mobile-health
deployment serving a specific user base, whereas $\alpha = 0.5$
yields a more balanced distribution approaching the IID
regime~\cite{yurochkin2019bayesian}. Under strong skew, a client
holding mainly \emph{neutral} and \emph{anger} samples produces
updates that are misaligned with those of a client holding
\emph{happiness} and \emph{sadness} samples, exacerbating client
drift and degrading the global Unweighted Average Recall (UAR).
}

{\color{black}
The second is \emph{resource heterogeneity}: clients run on
smartphones, tablets, and laptops with different CPU frequencies,
memory, and network conditions. Local computation is modelled on
CPUs only, consistent with deployment on personal mobile and edge
devices, to isolate the scheduling and selection
problem~\cite{nishio2019client, lai2021oort}; extending the model
to GPU-equipped devices is left for future work. For client $k$,
let $C_k$ denote the CPU cycles required per sample, $D_k$ the
number of local samples, $b_k$ the batch size, $e_k$ the number
of local epochs, and $f_k$ the effective CPU processing rate. The
number of local iterations and the computation time are
\begin{equation}
N_k = \frac{D_k}{b_k}\, e_k ,
\qquad
\tau_{k,\mathrm{comp}}^{(t)} = \frac{N_k b_k C_k}{f_k}.
\label{eq:computation_time}
\end{equation}

\begin{figure*}[t]
    \centering
    \includegraphics[width=0.8\linewidth]{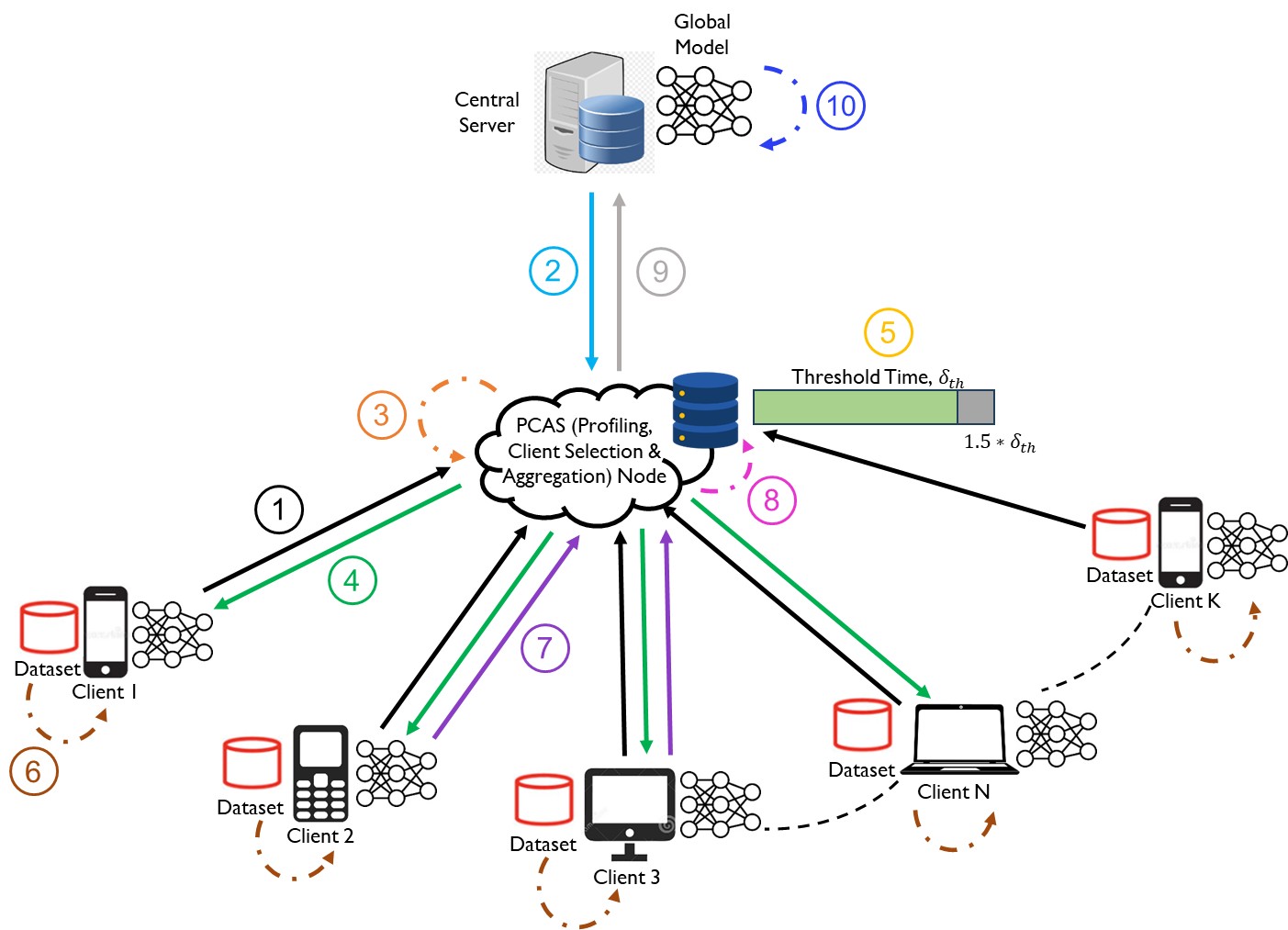}
    \caption{Server-Aware Client Selection Scheme for SER in Heterogeneous Federated Environments. $\scalebox{1.2}{\textcircled {\tiny{1}}}$ Clients sharing $O$ KPIs with PCAS. $\scalebox{1.2}{\textcircled {\tiny{2}}}$ Server sharing global model weights $\omega_{i,\text{glob}}^t$ along with $O$ CSWs with PCAS. $\scalebox{1.2}{\textcircled {\tiny{3}}}$ Ranking of $K$ clients by PCAS. $\scalebox{1.2}{\textcircled {\tiny{4}}}$ Selection and sharing of $\omega_{i,\text{glob}}^t$ with top-ranked $N$ clients by PCAS. $\scalebox{1.2}{\textcircled {\tiny{5}}}$ PCAS threshold timer to manage clients participation and stragglers. $\scalebox{1.2}{\textcircled {\tiny{6}}}$ Local training by $N$ clients using $\omega_{i,\text{glob}}^t$. $\scalebox{1.2}{\textcircled {\tiny{7}}}$ Sharing of updated weights $\omega_{i,k}^{t+1}$ and $O$ KPIs by $N$ clients with PCAS. $\scalebox{1.2}{\textcircled {\tiny{8}}}$ Update of global weights $\omega_{i,\text{glob}}^{t+1}$ by PCAS. $\scalebox{1.2}{\textcircled {\tiny{9}}}$ Sharing of updated global weights $\omega_{i,\text{glob}}^{t+1}$ with server for global model training. $\scalebox{1.2}{\textcircled {\tiny{10}}}$ Server updating global model with weights $\omega_{i,\text{glob}}^{t+1}$, obtaining immediate reward $R^t$ and updating quality values.}
    \label{fig:client_selection_scheme}
\end{figure*}

Communication time depends on the transferred model size and the
effective throughput of each link:
\begin{equation}
\tau_{k,\mathrm{comm}}^{(t)}
=
\frac{S_{\mathrm{down}}}{R_{k,\mathrm{down}}^{(t)}}
+
\frac{S_{\mathrm{up}}}{R_{k,\mathrm{up}}^{(t)}},
\label{eq:communication_time}
\end{equation}
where $S_{\mathrm{down}}$ and $S_{\mathrm{up}}$ are the sizes of
the downloaded global model and the uploaded local update, and
$R_{k,\mathrm{down}}^{(t)}$ and $R_{k,\mathrm{up}}^{(t)}$ are the
corresponding effective throughputs. Additional delays due to
propagation, queuing, protocol processing, and retransmissions
are captured by $\tau_{k,\mathrm{net}}^{(t)}$, giving the
end-to-end completion time
\begin{equation}
T_k^{(t)}
=
\tau_{k,\mathrm{comp}}^{(t)}
+
\tau_{k,\mathrm{comm}}^{(t)}
+
\tau_{k,\mathrm{net}}^{(t)} .
\label{eq:client_completion_time}
\end{equation}
The global round time is set by the slowest update included in
the aggregation:
\begin{equation}
T_{\mathrm{glob}}^{(t)}
=
\max_{k \in \mathcal{S}_{\mathrm{agg}}^{t}} T_k^{(t)},
\label{eq:global_round_time}
\end{equation}
where $\mathcal{S}_{\mathrm{agg}}^{t}$ is the set of clients whose
round-$t$ updates are aggregated in round $t$.

These latencies interact directly with client selection:
consistently selecting slow clients inflates cumulative training
time, whereas selecting only fast clients biases the global model
towards particular emotional distributions. This trade-off is
formalised next.
}

\subsubsection{Problem Formulation}
\label{subsec:problem}

{\color{black}
Let $\mathbf{x}^t = \{x_k^t\}_{k=1}^{K}$ denote the binary
selection variables at round $t$, where $x_k^t = 1$ if client $k$
is selected and $x_k^t = 0$ otherwise, and let
$\mathcal{S}^t = \{k : x_k^t = 1\}$ be the selected subset. Let
$Acc_{\mathrm{glob}}(\mathbf{x}^t)$ denote the UAR of the global
SER model after aggregating the updates of $\mathcal{S}^t$. The
objective is to maximise the global UAR under round-level timing
and participation constraints:
\begin{equation}
\max_{\mathbf{x}^t}
\quad
Acc_{\mathrm{glob}}(\mathbf{x}^t)
\label{eq:global_objective}
\end{equation}
subject to
\begin{equation}
\frac{T_{\mathrm{glob}}^{(t)}-\delta_{\mathrm{th}}}
{\delta_{\mathrm{th}}}
\leq \phi_T ,
\label{eq:timing_constraint}
\end{equation}
\begin{equation}
x_k^t \in \{0,1\},
\quad \forall k \in \{1,\ldots,K\},
\label{eq:binary_constraint}
\end{equation}
\begin{equation}
\sum_{k=1}^{K} x_k^t = N.
\label{eq:participation_constraint}
\end{equation}
Constraint~(\ref{eq:timing_constraint}) bounds the relative
deviation of the round time from the deadline
$\delta_{\mathrm{th}}$ by a tolerance $\phi_T$, preventing the
slowest client from dictating the round duration as in standard
FedAvg~\cite{mcmahan2017communication}.
Constraints~(\ref{eq:binary_constraint})
and~(\ref{eq:participation_constraint}) define the binary
participation decisions and fix the number of selected clients.

Solving this problem directly is impractical. Selecting $N$ of
$K$ clients yields $\binom{K}{N}$ candidate subsets per round,
making exhaustive search intractable for realistic
$K$~\cite{lai2021oort}. Moreover,
$Acc_{\mathrm{glob}}(\mathbf{x}^t)$ depends on the quality,
\divkpi{}, and timeliness of the selected updates through a
relationship that is non-stationary across rounds and cannot be
determined analytically in advance. An effective solution must
therefore learn its selection policy from round-level feedback.
}

\begin{table}[t!]
\centering
\caption{Salient Design Details of Adaptive Learning of CSWs by the Server Using Fuzzy Q-learning}
\label{tab:fuzzy_qlearning_pcas}
\renewcommand{\arraystretch}{0.75}
\setlength{\tabcolsep}{2pt}
\begin{tabular}{p{0.90\linewidth}}
\hline
\textbf{Aim} -
To learn optimal client selection weights corresponding to articulated fuzzy rules in a heterogeneous FL environment.  \\
\hline
\textbf{RL Agent} - The central server hosting and updating the global model. \\
\hline
\textbf{Fuzzified Inputs} - $Acc_{\text{glob}}$, $T_{\text{glob}}$ and $\zeta_{\text{glob}}$ \\
\multicolumn{1}{c}{%
\begin{minipage}{0.9\linewidth}
\centering
\includegraphics[width=\linewidth]{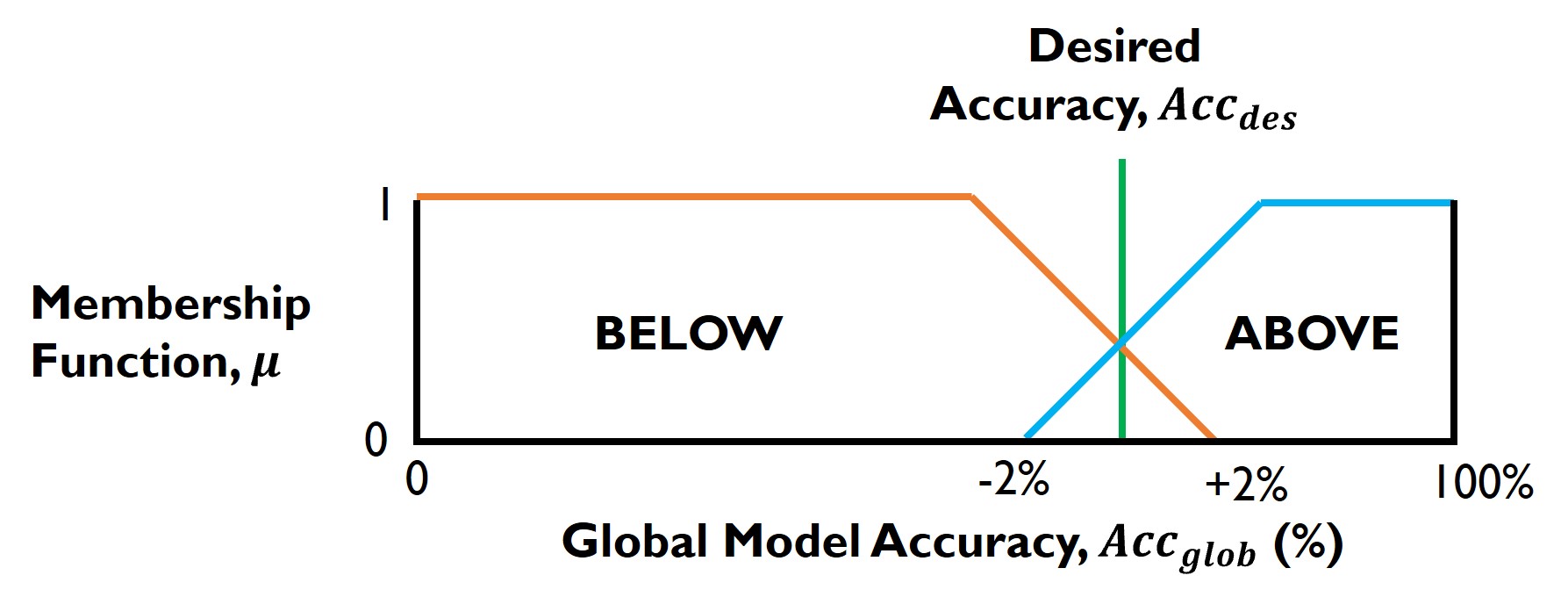}\par\vspace{0.1mm}
\includegraphics[width=\linewidth]{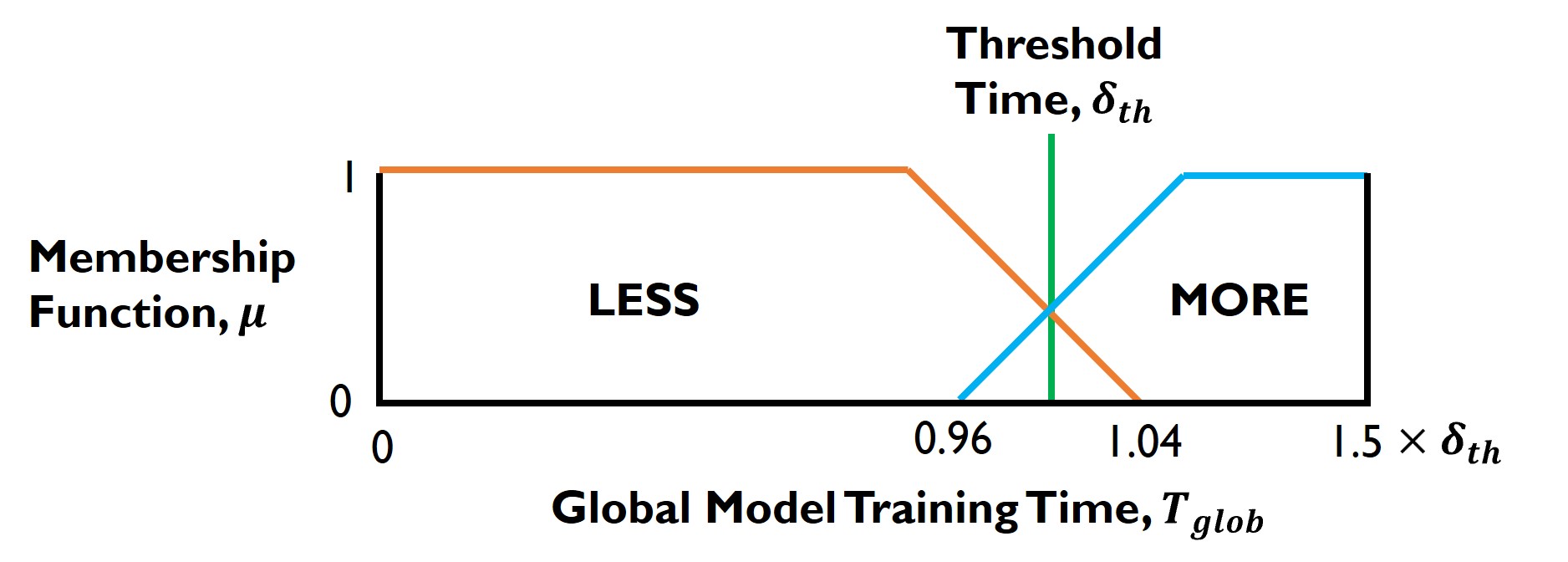}\par\vspace{0.1mm}
\includegraphics[width=\linewidth]{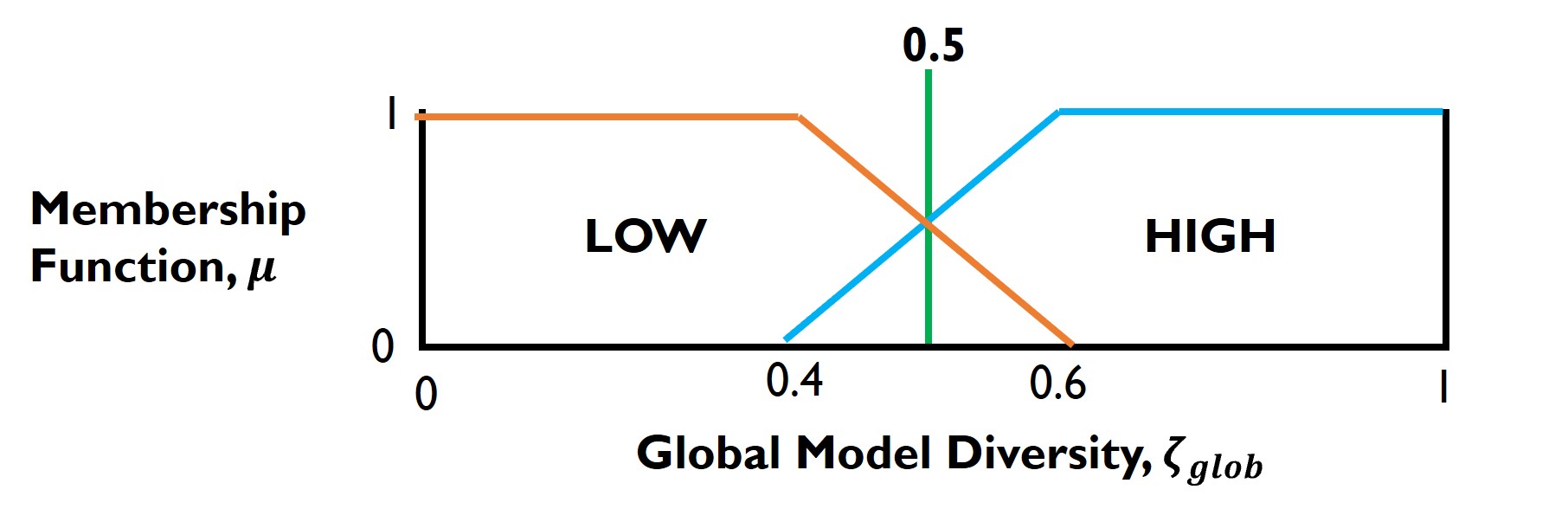}
\end{minipage}
} \\
\hline
\textbf{Fuzzy Rule Base} \\
\textbf{IF} $Acc_{\text{glob}}$ is \textbf{BELOW} AND $T_{\text{glob}}$ is \textbf{LESS} AND $\zeta_{\text{glob}}$ is \textbf{LOW} THEN \textbf{conc} is $\dots$ \\
\textbf{IF} $Acc_{\text{glob}}$ is \textbf{BELOW} AND $T_{\text{glob}}$ is \textbf{LESS} AND $\zeta_{\text{glob}}$ is \textbf{HIGH} THEN \textbf{conc} is $\dots$ \\
\textbf{IF} $Acc_{\text{glob}}$ is \textbf{BELOW} AND $T_{\text{glob}}$ is \textbf{MORE} AND $\zeta_{\text{glob}}$ is \textbf{LOW} THEN \textbf{conc} is $\dots$ \\
\textbf{IF} $Acc_{\text{glob}}$ is \textbf{BELOW} AND $T_{\text{glob}}$ is \textbf{MORE} AND $\zeta_{\text{glob}}$ is \textbf{HIGH} THEN \textbf{conc} is $\dots$ \\
\textbf{IF} $Acc_{\text{glob}}$ is \textbf{ABOVE} AND $T_{\text{glob}}$ is \textbf{LESS} AND $\zeta_{\text{glob}}$ is \textbf{LOW} THEN \textbf{conc} is $\dots$ \\
\textbf{IF} $Acc_{\text{glob}}$ is \textbf{ABOVE} AND $T_{\text{glob}}$ is \textbf{LESS} AND $\zeta_{\text{glob}}$ is \textbf{HIGH} THEN \textbf{conc} is $\dots$ \\
\textbf{IF} $Acc_{\text{glob}}$ is \textbf{ABOVE} AND $T_{\text{glob}}$ is \textbf{MORE} AND $\zeta_{\text{glob}}$ is \textbf{LOW} THEN \textbf{conc} is $\dots$ \\
\textbf{IF} $Acc_{\text{glob}}$ is \textbf{ABOVE} AND $T_{\text{glob}}$ is \textbf{MORE} AND $\zeta_{\text{glob}}$ is \textbf{HIGH} THEN \textbf{conc} is $\dots$
 \\
\hline
\textbf{Conclusions} - The set of possible CSW values represented by the vector $\mathbf{a} = \{0,3,6\}$\\
\hline
\textbf{Optimisation Parameters} - CSWs representing KPIs ($Acc_{\text{glob}}$, $T_{\text{glob}}$ and $\zeta_{\text{glob}}$) of the global model \\
\hline
\textbf{Immediate Reward} - The immediate reward $R^t$ is computed in accordance with \eqref{eq:reward} \\
\hline
\textbf{Cumulative Rule-Conclusion Quality Value} - Cumulative quality value for each fuzzy rule-conclusion pair $q(\phi_i,a_i)$ is updated as \eqref{eq:q_update} \\
\hline
\end{tabular}
\end{table}

\subsection{Proposed Solution}
\label{subsec:solution}

{\color{black}
\textit{Fuzzy-FedAvg} solves this problem through a three-party
architecture, shown in Figure~\ref{fig:client_selection_scheme}.
The server maintains the global model and learns, through fuzzy
Q-learning, the CSWs that set the relative importance of accuracy,
training time, and diversity. The PCAS node translates these
weights into client-ranking and participation decisions, manages
the submission deadlines, and aggregates the received updates.
Although these client-facing operations could be performed by the
server itself, assigning them to PCAS separates client
coordination and straggler management from the learning
policy~\cite{mcmahan2017communication, bonawitz2019towards}. The
selection policy can therefore be modified without changing the
coordination mechanism, and coordination responsibilities can be
distributed in larger deployments.

Each round forms a closed loop of three mechanisms. The client
selection mechanism (Section~\ref{subsubsec:selection}) decides
\emph{which} clients participate; the learning mechanism
(Section~\ref{subsubsec:learning}) decides \emph{how} the server
adapts its priorities from the observed reward; and the
aggregation mechanism (Section~\ref{subsubsec:aggregation})
combines the updates using the same priorities.
Table~\ref{tab:fuzzy_qlearning_pcas} summarises the design of the
learning mechanism.
}

\subsubsection{Client Selection Mechanism}
\label{subsubsec:selection}

{\color{black}
At the start of each round, all $K$ clients report three KPIs to
the PCAS node, each normalised to a scale of 1--5 (5 = best):
local accuracy $Acc_k^t$, training time $T_k^t$, and local model
diversity $\zeta_k^t$. This adds only $3K$ scalars per round,
negligible compared with model exchange~\cite{bonawitz2019towards}.
Accuracy reflects the quality of a client's update, and training
time reflects its resource efficiency and straggler tendency.
Diversity is measured as the cosine distance between the locally
updated and the received global model parameters:
\begin{equation}
\zeta_k^{t+1}
=
\min\left\{
1,\,
\max\left[
0,\,
1-
\frac{
\boldsymbol{\omega}_{i,k}^{t+1}
\cdot
\boldsymbol{\omega}_{i,\mathrm{glob}}^{t}
}{
\left\|\boldsymbol{\omega}_{i,k}^{t+1}\right\|
\left\|\boldsymbol{\omega}_{i,\mathrm{glob}}^{t}\right\|
}
\right]
\right\},
\label{eq:model_diversity}
\end{equation}
where $\zeta_k = 0$ indicates aligned parameter vectors and
$\zeta_k = 1$ indicates orthogonal or oppositely directed vectors
after clipping. The score measures the distinctiveness of a
client's update rather than its data distribution directly,
although clients with more heterogeneous data tend to produce more
divergent updates~\cite{li2020fedprox}. Favouring such clients
brings in information from under-explored regions of the loss
landscape and counteracts the repeated selection of similar
clients~\cite{lai2021oort}. Treating diversity as a selection
criterion in its own right extends the utility--efficiency view of
Oort~\cite{lai2021oort} and the loss-based view of
Power-of-Choice~\cite{cho2022client}.

Rather than optimising $\mathbf{x}^t$ directly, the PCAS node
ranks clients using the CSWs
$\boldsymbol{\nu}^{\,t}=\{\nu_T^t,\nu_{\mathrm{Acc}}^t,\nu_{\zeta}^t\}$
learned by the server:
\begin{equation}
r_k^t = \sum_{q=1}^{O} \mu_{q,k}^t \, \nu_q^t, \quad \forall k \in \{1,\ldots,K\},
\label{eq:client_ranking}
\end{equation}
where $\mu_{q,k}^t$ is the normalised value of the $q^{th}$ KPI of
client $k$ and $\nu_q^t$ is its CSW. The $N$ highest-ranked
eligible clients are then selected:
\begin{equation}
x_k^t =
\begin{cases}
1, & \text{if } k \in
\operatorname{TopN}\!\left(\{r_j^t\}_{j=1}^{K}\right),\\
0, & \text{otherwise},
\end{cases}
\label{eq:binary_selection}
\end{equation}
which reparameterises the combinatorial problem
in~(\ref{eq:global_objective})--(\ref{eq:participation_constraint})
over the continuous CSWs:
\begin{equation}
\max_{\boldsymbol{\nu}^{\,t}\in\mathcal{A}}
\quad
Acc_{\mathrm{glob}}\!\left(\mathbf{x}^t(\boldsymbol{\nu}^{\,t})\right),
\label{eq:csw_objective}
\end{equation}
subject to the same constraints, where $\mathcal{A}$ is the set of
admissible CSW configurations defined by the fuzzy Q-learning
action space. Constraint~(\ref{eq:participation_constraint}) is
satisfied by construction.

The selected clients receive the current global model
$\omega_{i,\text{glob}}^t$. To avoid repeatedly selecting the same
subset, which would under-represent minority emotional
distributions, clients that submit an update are excluded from
selection for the next three rounds. This cooldown was set
empirically: shorter periods did not break repeated selection
under $\alpha = 0.1$, while longer ones unnecessarily excluded
high-quality clients.

Because of resource heterogeneity, the selected clients do not
finish at the same time. Rather than waiting for the slowest
client, dropping stragglers~\cite{mcmahan2017communication},
accepting the staleness of fully asynchronous
FL~\cite{xie2019asynchronous}, or pre-profiling clients into
latency tiers~\cite{chai2020tifl}, the PCAS node classifies
clients by their completion time
$T_k^{(t)}$ in~(\ref{eq:client_completion_time}) relative to a
deadline $\delta_{\mathrm{th}}$. \textbf{Tier~1} (\textit{timely})
clients, finishing within $1.25\times\delta_{\mathrm{th}}$, join
the current aggregation. \textbf{Tier~2} (\textit{deferred})
clients, finishing between $1.25\times$ and
$1.5\times\delta_{\mathrm{th}}$, are carried over to the next
round rather than discarded. \textbf{Tier~3} (\textit{straggler})
clients, exceeding $1.5\times\delta_{\mathrm{th}}$, are excluded
from selection for two rounds. The first 25\% margin absorbs
minor latency fluctuations, and the second bounds the age of
deferred updates to one round.

Deferred updates are attenuated by a staleness factor. With
$s_k^t = t - t_k^{\mathrm{last}}$, where $t_k^{\mathrm{last}}$ is
the round in which the update was generated,
\begin{equation}
\lambda_k^t = \max\left(0.5,\,1-0.1s_k^t\right).
\label{eq:staleness_factor}
\end{equation}
A fresh update thus receives $\lambda_k^t=1$, and a Tier-2 update,
whose staleness is bounded to one round, receives
$\lambda_k^t=0.9$. Applying $\lambda_k^t$ in the aggregation
weights~(\ref{eq:per_client_weights}) retains completed
computation while reducing the influence of stale updates.

Both the ranking and, later, the aggregation depend on the CSWs.
How the server learns them is described next.
}

\subsubsection{Learning Mechanism}
\label{subsubsec:learning}

{\color{black}
Fixed CSWs cannot track how the importance of accuracy, diversity,
and training time changes during training: what helps in early
rounds, when the model is far from convergence, differs from what
helps in later rounds, when gains are marginal and stability
matters. The server therefore learns the CSWs through fuzzy
Q-learning~\cite{berenji1992learning, pervez2017fuzzy}, which
combines fuzzy inference with Q-learning so that the agent can
generalise over the continuous, noisy KPI space without a large
discrete Q-table.} Following the Mamdani fuzzy inference
framework~\cite{mamdani1975experiment}, each rule fires
with a degree of truth equal to the minimum membership
value across its antecedents; fired conclusions are
aggregated via union and defuzzified via centroid to
produce crisp CSW values.

{\color{black}
The server's state at round $t$ is the global KPI tuple
$(Acc_{\text{glob}}^t,\, T_{\text{glob}}^t,\, \zeta_{\text{glob}}^t)$.
Each KPI is fuzzified into two linguistic values
(Table~\ref{tab:fuzzy_qlearning_pcas}): $Acc_{\text{glob}}$ into
\textbf{ABOVE}/\textbf{BELOW} relative to the desired accuracy
$Acc_{des}$, $T_{\text{glob}}$ into \textbf{MORE}/\textbf{LESS}
relative to $\delta_{\text{th}}$, and $\zeta_{\text{glob}}$ into
\textbf{HIGH}/\textbf{LOW} relative to a diversity threshold. This
gives $2^3 = 8$ rules, compact enough to converge within typical
FL training horizons~\cite{berenji1992learning, pervez2017fuzzy}.
The rules encode intuitive responses: when accuracy and diversity
are both low, raising the diversity CSW draws in clients from
under-represented emotional distributions; when rounds are slow,
raising the time CSW restores throughput; and when all KPIs are
favourable, prioritising accuracy consolidates the gains.

Each activated rule $\phi_i$ selects a conclusion $a_i$ from
$\mathbf{a} = \{0,\, 3,\, 6\}$, corresponding to ignoring,
moderately prioritising, or highly prioritising the KPI. The agent
explores the rule--conclusion space during the early rounds and
then alternates based on the reward: it continues exploring while
$R^t$ is positive and exploits the best-known conclusions
otherwise. The inferred CSW for the current fuzzy state $s_f$ is
\begin{equation}
H(s_f) = \frac{\sum_{i=1}^{m} w_i a_i}{\sum_{i=1}^{m} w_i},
\label{eq:inferred_action}
\end{equation}
where $w_i$ is the firing strength of the $i^{th}$ activated rule.
After each round, the quality value of each rule--conclusion pair
is updated as
\begin{equation}
q(\phi^t_i, a^t_i) \leftarrow q(\phi^t_i, a^t_i) + \eta_Q\, \Delta q(\phi^t_i, a^t_i),
\label{eq:q_update}
\end{equation}
\begin{equation}
\begin{aligned}
\Delta q(\phi_i, a_i) &=
\frac{w_i}{\sum_{j=1}^{m} w_j}
\Big[
R^t + \gamma_Q\, FQ^*(s_f', H(s_f')) \\
&\quad - FQ(s_f, H(s_f))
\Big],
\end{aligned}
\label{eq:deltaq}
\end{equation}
where $\eta_Q \in (0,1)$ is the learning rate,
$\gamma_Q \in [0,1)$ is the discount factor, and the fuzzy
Q-values of the current state $s_f$ and the next state $s_f'$ are
\begin{equation}
FQ(s_f, H(s_f)) =
\frac{\sum_{i=1}^{m} w_i\, q(\phi_i, a_i)}{\sum_{i=1}^{m} w_i},
\label{eq:fq}
\end{equation}
\begin{equation}
FQ^*(s_f', H(s_f')) =
\frac{\sum_{i=1}^{m} w_i'\, q(\phi_i, a_i^*)}{\sum_{i=1}^{m} w_i'},
\label{eq:fq_star}
\end{equation}
with $w_i'$ the firing strengths in the next state and
$a_i^* = \arg\max_{a_j \in \mathbf{a}} q(\phi_i, a_j)$.

The reward $R^t$ combines four terms and a penalty, each targeting
a distinct failure mode that a single scalar cannot
distinguish~\cite{sutton2018reinforcement}:
\begin{equation}
R^t = R^t_p + R^t_b + R^t_g + R^t_m - S^t_p .
\label{eq:reward}
\end{equation}
The \textbf{progress reward}
$R^t_p = 200(Acc_{\text{glob}}^t - Acc_{\text{glob}}^{t-1})$
favours steady improvement and dominates the signal, so that even
small gains are rewarded. The \textbf{milestone reward}
$R^t_b = 150(Acc_{\text{glob}}^t - Acc_{\text{best}})$
encourages surpassing the best model so far, discouraging
premature convergence~\cite{li2020fedprox}. The \textbf{gap
reward}
\begin{equation*}
R^t_g =
\begin{cases}
30(0.2 - \Delta_{\text{gap}}^t), & \text{if } \Delta_{\text{gap}}^t < 0.2,\\
0, & \text{otherwise},
\end{cases}
\end{equation*}
with $\Delta_{\text{gap}}^t = Acc_{\text{cent}} - Acc_{\text{glob}}^t$,
activates only once the federated model is within 20\% of
centralised accuracy. The \textbf{threshold reward}
\begin{equation*}
R^t_m =
\begin{cases}
80, & \text{if } Acc_{\text{glob}}^t \ge Acc_m,\\
0, & \text{otherwise},
\end{cases}
\end{equation*}
prevents the agent from trading accuracy for speed. Finally, the
\textbf{stagnation penalty}
\begin{equation*}
S_p^t =
\begin{cases}
\min(15, S_c), & \text{if } Acc_{\text{glob}}^t - Acc_{\text{glob}}^{t-1} < \varepsilon_s,\\
0, & \text{otherwise},
\end{cases}
\end{equation*}
where $S_c$ is the stagnation count and $\varepsilon_s$ is the
minimum improvement threshold, discourages persisting with a
failing strategy, with the cap of 15 avoiding over-penalisation
during unavoidable plateaus.

The CSWs learned in this way are used not only to select clients
but also to aggregate their updates.
}

\subsubsection{Aggregation Mechanism}
\label{subsubsec:aggregation}

{\color{black}
The CSWs that rank clients in~(\ref{eq:client_ranking}) also
weight their updates during aggregation, so that selection and
aggregation follow the same adaptive strategy. Let
$\mathcal{U}^t$ denote the updates aggregated at round $t$,
i.e., the current Tier-1 updates and the Tier-2 updates deferred
from the previous round. For KPI
$q \in \{T, \mathrm{Acc}, \zeta\}$, the normalised weight of
client $k \in \mathcal{U}^t$ is
\begin{equation}
\tilde{w}_{q,k} =
\frac{\lambda_k^t\, g_q(k)}
{\sum_{j\in\mathcal{U}^t} \lambda_j^t\, g_q(j)},
\label{eq:per_client_weights}
\end{equation}
where $g_T(k) = 1/T_k$, $g_{\mathrm{Acc}}(k) = Acc_k$,
$g_{\zeta}(k) = \zeta_k$, and $\lambda_k^t$ is the staleness
factor in~(\ref{eq:staleness_factor}). The global model is
updated as
\begin{equation}
\omega_{i,\mathrm{glob}}^{\,t+1} =
\frac{1}{\hat{\nu}} \sum_{q} \nu_q^t
\sum_{k\in\mathcal{U}^t} \tilde{w}_{q,k}\,
\omega_{i,k}^{\,t+1},
\label{eq:aggregation}
\end{equation}
where $\hat{\nu} = \sum_{q} \nu_q^t$ keeps the update a convex
combination of local models. Faster, more accurate, and more
diverse clients therefore contribute more strongly, while
deferred updates are attenuated by $\lambda_k^t$ rather than
discarded. This coupling of selection and aggregation through
shared CSWs distinguishes \textit{Fuzzy-FedAvg} from prior
methods that treat the two steps
independently~\cite{lai2021oort, cho2022client}.

Algorithm~\ref{alg:fuzzy_fedavg} summarises the complete
procedure, with each step corresponding to the annotated flow in
Figure~\ref{fig:client_selection_scheme}. The fuzzy Q-learning
agent adds $\mathcal{O}(1)$ overhead per round (fuzzification,
evaluation of eight rules, defuzzification, and one Q-table
update), independent of $K$ and $G$.
}

\begin{algorithm}[!ht]
\caption{\textit{Fuzzy-FedAvg}: Fuzzy Q-learning-based
         Server-Aware Client Selection for Federated SER}
\label{alg:fuzzy_fedavg}
\begin{algorithmic}[1]

\Require $K$, $N$, $G$, $e_k$, $b_k$, $\delta_{\text{th}}$,
         $\phi_T$, $Acc_{\text{des}}$, $Acc_m$,
         {\color{black}$\eta_Q$}, $\gamma_Q$, $\varepsilon$,
         $\varepsilon_s$
\Ensure  $\omega_{i,\text{glob}}^{G}$

\State Initialise $\omega_{i,\text{glob}}^{0}$ randomly;
       $\mathcal{Q} \leftarrow \mathbf{0}$;
       $Acc_{\text{best}} \leftarrow 0$;
       $S_c \leftarrow 0$;
       $\mathcal{B} \leftarrow \emptyset$

\For{$t = 1, 2, \dots, G$}

  \Statex \quad \textit{--- \ding{192} Clients share KPIs with PCAS ---}
  \For{$k = 1$ \textbf{to} $K$} \Comment{in parallel}
    \State Compute $(Acc_k^t, T_k^t, \zeta_k^t)$;
           normalise to $[1,5]$; transmit to PCAS
  \EndFor

  \Statex \quad \textit{--- \ding{193} Server shares weights + CSWs ---}
  \State Observe $s^t \!=\!
         (Acc_{\text{glob}}^{t-1},
          T_{\text{glob}}^{t-1},
          \zeta_{\text{glob}}^{t-1})$;
         fuzzify per Table~\ref{tab:fuzzy_qlearning_pcas}
  \State Select $\mathbf{a}^t$ via
         proposed \textit{exploration-exploitation} scheme on $\mathcal{Q}$;
         compute $\hat{\nu}$; transmit
         $\omega_{i,\text{glob}}^{t-1}$ and CSWs to PCAS

  \Statex \quad \textit{--- \ding{194} PCAS ranks and selects clients ---}
  \State {\color{black}$r_k^t \leftarrow
         \sum_{q} \mu_{q,k}^t \, \nu_q^t$},
         $\forall\, k$; select top-$N$ (exclude 3-round
         recent participants and penalised Tier-3 clients)

  \Statex \quad \textit{--- \ding{195}\ding{196} PCAS distributes model; starts timer ---}
  \State Distribute $\omega_{i,\text{glob}}^{t-1}$
         to selected clients; start $\delta_{\text{th}}$

  \Statex \quad \textit{--- \ding{197} Clients perform local training ---}
  \For{$n = 1$ \textbf{to} $N$} \Comment{in parallel}
    \State Store copy of $\omega_{i,\text{glob}}^{t-1}$;
           train $e_n$~epochs, batch $b_n$
    \State Compute $\zeta_n^{t+1}$ via
           {\color{black}Eq.~\eqref{eq:model_diversity}}
  \EndFor

  \Statex \quad \textit{--- \ding{198} Clients share weights + KPIs ---}
  \For{$n = 1$ \textbf{to} $N$} \Comment{in parallel}
    \State Transmit
           $(\omega_{i,n}^{t+1}, Acc_n^{t+1},
             T_n^{t+1}, \zeta_n^{t+1})$ to PCAS
  \EndFor

  \Statex \quad \textit{--- \ding{199} PCAS classifies clients and aggregates ---}
  \State Tier~1~(${\le}1.25\delta_{\text{th}}$): current round;
         Tier~2~($(1.25\text{--}1.5)\delta_{\text{th}}$): defer to
         $\mathcal{B}$; Tier~3~(${{>}}1.5\delta_{\text{th}}$):
         2-round ban
  \State Aggregate $\omega_{i,\text{glob}}^{t+1}$ via
         Eq.~\eqref{eq:aggregation} using
         {\color{black}$\mathcal{U}^t =$ Tier-1 $\cup\,\mathcal{B}$ with $\lambda_k^t$ from Eq.~\eqref{eq:staleness_factor}};
         update $\mathcal{B}$

  \Statex \quad \textit{--- \ding{200} PCAS shares weights ---}
  \State Transmit $\omega_{i,\text{glob}}^{t+1}$ to server

  \Statex \quad \textit{--- \ding{201} Server updates global model ---}
  \State Compute $R^t$ via Eq.~\eqref{eq:reward}
  \State Update $q(\phi_i, a_i)$ using Eq.~\eqref{eq:q_update}
  \State $Acc_{\text{best}} \leftarrow
         \max(Acc_{\text{best}},
              Acc_{\text{glob}}^{t+1})$;
         update $S_c$; decay $\varepsilon$

  \If{$Acc_{\text{glob}}^{t+1} \geq Acc_{\text{des}}$}
    \State \textbf{break}
  \EndIf

\EndFor
\State \Return $\omega_{i,\text{glob}}^{G}$

\end{algorithmic}
\end{algorithm}
\section{Experimental Setup}
\label{sec:experimental_setup}

\subsection{Datasets and Evaluation Protocol}
{\color{black}The proposed framework is evaluated on two widely used SER benchmarks, IEMOCAP~\cite{busso2008iemocap} and CREMA-D~\cite{cao2014crema}. For IEMOCAP, only the improvised utterances corresponding to four emotion classes --- happy, sad, angry, and neutral --- are considered. To prevent speaker and session leakage, a strictly session-disjoint protocol is adopted. Sessions~1--3 are used exclusively to construct the federated training-client pool, Session~4 is reserved for validation and round-wise global-model evaluation, and Session~5 is held out and accessed only for final testing. Consequently, no utterance or speaker appearing in the final test session contributes to client training or model selection.

For CREMA-D, the six emotion classes happy, sad, angry, neutral, fear, and disgust are considered. A speaker-disjoint split is used, in which approximately 70\% of the speakers are assigned to training, 15\% to validation, and the remaining speakers to the final test set. The speaker split is fixed across methods and experimental seeds. Only utterances belonging to training speakers are subsequently partitioned among federated clients. Thus, speakers in the validation and test partitions are never included in any client's local training data.}

\subsection{Acoustic Features and SER Model}

{\color{black}Two acoustic representations are evaluated: Mel-frequency cepstral coefficients (MFCCs) and Mel-spectrograms. All utterances are resampled to 16~kHz and framed using a 25~ms Hamming window with a 10~ms hop. For the MFCC representation, 40 coefficients are extracted per frame, capturing the spectral envelope of speech in a compact, decorrelated form. For the Mel-spectrogram representation, 128 Mel filter banks are applied and the resulting energies are converted to a logarithmic (decibel) scale, preserving richer time--frequency detail. Utterances are zero-padded or truncated to a fixed number of frames so that each sample yields a fixed-size two-dimensional input. The same feature representation and model architecture are used for all federated methods within a given experimental configuration. All federated methods use the same convolutional SER model to ensure a controlled comparison. The model consists of three convolutional blocks with 64, 128, and 256 filters, respectively. Each block contains a $5\times5$ convolutional layer followed by batch normalisation and $2\times2$ max pooling. The convolutional backbone is followed by global average pooling, a fully connected layer with 512 units, dropout, and a softmax output layer. ReLU activation is used in the hidden layers. The model is trained using Adam with a learning rate of $10^{-4}$ and categorical cross-entropy loss.}

\subsection{Federated Data Partitioning and System Configuration}

{\color{black}The training partition is distributed across $K=30$ simulated clients using the Dirichlet partitioning in~(\ref{eq:dirichlet}) with concentration parameter $\alpha$. Smaller values of $\alpha$ produce more class-skewed local datasets and therefore stronger statistical heterogeneity. The experiments consider $\alpha\in\{0.1,0.3,0.5,1.0\}$, with $N\in\{4,8,16\}$ clients participating in each communication round and training horizons of $G\in\{40,60,80,120\}$ rounds, depending on the experiment. The reference configuration consists of $K=30$, $N=8$, $\alpha=0.1$, and $G=80$; parameters varied in the sensitivity experiments are reported in the corresponding results subsections.

System heterogeneity is represented using three client-device profiles corresponding to low-, medium-, and high-capability devices (smartphones, tablets, and laptops), which differ in computational capacity, local epoch count, and batch size. For the dedicated timing evaluation, heterogeneous execution profiles are derived from FedScale to provide a more realistic representation of client-side system variability. In this evaluation, client completion time follows~(\ref{eq:client_completion_time}): the computation component reflects device-dependent processing capability, while the communication component is represented through effective uplink/downlink throughput and additional network latency, rather than a single synthetic delay term.

The \textit{Fuzzy-FedAvg}-specific settings follow Section~\ref{subsec:solution}: CSW conclusions are drawn from $\mathbf{a}=\{0,3,6\}$; clients are classified as timely, deferred, or stragglers at $1.25\times$ and $1.5\times\delta_{\mathrm{th}}$; participating clients observe a three-round selection cooldown; stragglers are excluded for two rounds; and deferred updates are attenuated according to~(\ref{eq:staleness_factor}). Unless otherwise stated, all compared methods use identical client partitions, model initialisation, model architecture, validation/test partitions, participation budget, and training horizon within each experimental configuration. The principal simulation parameters are summarised in Table~\ref{tab:simulation_parameters}.}

\begin{table}[t!]
\centering
\caption{Simulation parameters used in the experimental evaluation.}
\label{tab:simulation_parameters}
\setlength{\tabcolsep}{4pt}
\begin{tabular}{p{2.7cm} p{4.2cm}}
\hline
\textbf{Parameter} & \textbf{Value / Description} \\ \hline
Datasets & IEMOCAP, CREMA-D \\ \hline
Acoustic Features & \rv{40 MFCCs; 128-band log-Mel-spectrogram (16~kHz, 25~ms window, 10~ms hop)} \\ \hline
Number of Clients, $K$ & 30 \\ \hline
Clients per Round, $N$ & \rv{8, 10, 12} \\ \hline
Dirichlet Parameter, $\alpha$ & 0.1, 0.3, 0.5, 1.0 \\ \hline
Global Rounds, $G$ & \rv{60, 80, 120} \\ \hline
Client Devices & Smartphone, Tablet, Laptop \\ \hline
Device Capabilities &
\textbf{Low:} comp. power $=1\times10^{6}$, epochs $=2$, batch size $=32$ \newline
\textbf{Medium:} comp. power $=2\times10^{6}$, epochs $=4$, batch size $=64$ \newline
\textbf{High:} comp. power $=4\times10^{6}$, epochs $=8$, batch size $=128$ \\ \hline
\rv{Round Deadline, $\delta_{\mathrm{th}}$} & 8 seconds \\ \hline
\rv{Tier Thresholds} & \rv{$1.25\times$ and $1.5\times\delta_{\mathrm{th}}$} \\ \hline
\rv{Selection Cooldown / Straggler Exclusion} & \rv{3 rounds / 2 rounds} \\ \hline
\rv{Staleness Factor, $\lambda_k^t$} & \rv{$\max(0.5,\,1-0.1s_k^t)$} \\ \hline
\rv{CSW Conclusion Set, $\mathbf{a}$} & \rv{$\{0,3,6\}$} \\ \hline
Desired Accuracy, $\mathrm{Acc}_{\mathrm{des}}$ & 0.6 \\ \hline
Fuzzy Q-learning Discount Factor\rv{, $\gamma_Q$} & 0.9 \\ \hline
Fuzzy Q-learning Learning Rate\rv{, $\eta_Q$} & $1\times10^{-5}$ \\ \hline
Random Seeds & 42, 137, 911 \\ \hline
Reference Configuration & $K=30$, $N=8$, $\alpha=0.1$, $G=80$ \\ \hline
\end{tabular}
\end{table}

\subsection{Evaluation Metrics and Statistical Reporting}

{\color{black}The principal SER performance metric is Unweighted Average Recall (UAR), the mean of per-class recalls, which weights all emotion classes equally and is therefore preferred over accuracy in the class-imbalanced SER setting. Accuracy is additionally reported where relevant. System efficiency is evaluated using cumulative training time, while client-level performance balance is assessed using participation fairness, accuracy fairness, accuracy disparity, and worst-case client accuracy. To account for stochastic variability, all experiments are repeated using three independent random seeds $\{42,137,911\}$. Results are reported as mean $\pm$ standard deviation across the three runs unless explicitly stated otherwise. The same seed is used for all competing methods within each paired experimental configuration.}

\begin{table*}[t]
\centering
\caption{Comprehensive comparison with stronger federated learning and
client-selection baselines under severe statistical heterogeneity
($\alpha=0.1$), with $K=30$, $N=8$, and $G=80$, using MFCC features.
Results are reported as mean $\pm$ standard deviation over three independent
runs using seeds 42, 137, and 911. Cumulative training time is obtained
under the common simulation timing setup used for all methods.
$\uparrow$ indicates higher is better and $\downarrow$ indicates lower is better. PF (Participation Fairness), AF (Accuracy Fairness), AD (Accuracy Disparity), WCA (Worst Case Accuracy).}
\label{tab:strong_baselines_comprehensive}
\scriptsize
\setlength{\tabcolsep}{3.5pt}
\renewcommand{\arraystretch}{1.15}
{\color{black}
\begin{tabular}{llcccccc}
\hline
\textbf{Dataset} & \textbf{Method} & \textbf{UAR $\uparrow$} & \textbf{Time (s) $\downarrow$} & \textbf{PF $\uparrow$} & \textbf{AF $\uparrow$} & \textbf{AD $\downarrow$} & \textbf{WCA $\uparrow$} \\
\hline
\multirow{6}{*}{IEMOCAP}
& FedAvg & $0.394 \pm 0.084$ & $669.05 \pm 127.59$ & $0.964 \pm 0.005$ & $0.929 \pm 0.009$ & $0.704 \pm 0.085$ & $0.241 \pm 0.077$ \\
& FedProx & $0.412 \pm 0.044$ & $481.94 \pm 82.47$ & $\mathbf{0.967 \pm 0.006}$ & $0.914 \pm 0.018$ & $0.739 \pm 0.082$ & $0.215 \pm 0.079$ \\
& SCAFFOLD & $0.501 \pm 0.029$ & $432.49 \pm 80.45$ & $\mathbf{0.967 \pm 0.006}$ & $0.935 \pm 0.006$ & $0.617 \pm 0.030$ & $0.209 \pm 0.067$ \\
& Power-of-Choice & $\mathbf{0.503 \pm 0.038}$ & $\mathbf{373.98 \pm 8.31}$ & $0.562 \pm 0.044$ & $\mathbf{0.959 \pm 0.008}$ & $\mathbf{0.549 \pm 0.050}$ & $\mathbf{0.386 \pm 0.064}$ \\
& Oort & $0.365 \pm 0.042$ & $467.89 \pm 127.56$ & $0.889 \pm 0.017$ & $0.928 \pm 0.029$ & $0.699 \pm 0.090$ & $0.238 \pm 0.078$ \\
& Fuzzy-FedAvg & $0.437 \pm 0.060$ & $505.14 \pm 25.19$ & $0.935 \pm 0.013$ & $0.915 \pm 0.007$ & $0.732 \pm 0.067$ & $0.191 \pm 0.063$ \\
\hline
\multirow{6}{*}{CREMA-D}
& FedAvg & $0.346 \pm 0.077$ & $1236.48 \pm 390.84$ & $\mathbf{0.967 \pm 0.010}$ & $0.874 \pm 0.024$ & $0.646 \pm 0.064$ & $0.187 \pm 0.023$ \\
& FedProx & $0.403 \pm 0.039$ & $854.38 \pm 212.87$ & $0.959 \pm 0.017$ & $\mathbf{0.901 \pm 0.026}$ & $0.589 \pm 0.027$ & $\mathbf{0.241 \pm 0.072}$ \\
& SCAFFOLD & $\mathbf{0.479 \pm 0.012}$ & $782.66 \pm 221.47$ & $0.959 \pm 0.017$ & $0.895 \pm 0.017$ & $\mathbf{0.537 \pm 0.034}$ & $0.094 \pm 0.024$ \\
& Power-of-Choice & $0.413 \pm 0.032$ & $882.67 \pm 294.30$ & $0.736 \pm 0.035$ & $0.868 \pm 0.059$ & $0.761 \pm 0.031$ & $0.128 \pm 0.039$ \\
& Oort & $0.378 \pm 0.056$ & $845.46 \pm 244.59$ & $0.869 \pm 0.024$ & $0.888 \pm 0.040$ & $0.684 \pm 0.097$ & $0.164 \pm 0.027$ \\
& Fuzzy-FedAvg & $0.418 \pm 0.029$ & $\mathbf{615.89 \pm 30.93}$ & $0.940 \pm 0.013$ & $0.893 \pm 0.024$ & $0.641 \pm 0.101$ & $0.174 \pm 0.020$ \\
\hline
\end{tabular}
}
\end{table*}

\section{Results}
\label{sec:results}

{\color{black}
This section first compares \textit{Fuzzy-FedAvg} with representative
federated-learning baselines and then examines its behaviour under
varying statistical heterogeneity, client participation, and training
horizons, followed by timing validation, ablation, and sensitivity
analyses. Unless otherwise stated, results are reported as mean $\pm$
standard deviation over seeds 42, 137, and 911 in the reference
configuration ($K=30$, $N=8$, $\alpha=0.1$, $G=80$).
}
\subsection{Comparison with Baselines and Client-Level Fairness}
\label{subsec:baselines}
\begin{figure}[!t]
\centering
\includegraphics[width=\columnwidth]{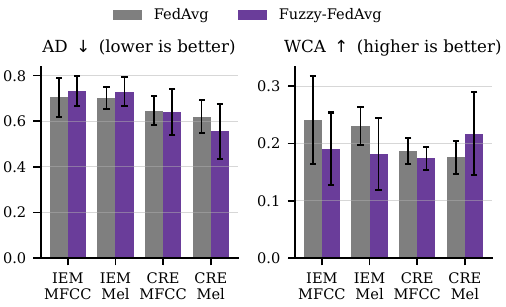}
\caption{\rv{Client-level fairness of FedAvg and \textit{Fuzzy-FedAvg} under severe statistical heterogeneity ($\alpha=0.1$, $K=30$, $N=8$, $G=80$): accuracy disparity (AD, lower is better) and worst-case accuracy (WCA, higher is better). Bars show the mean and error bars the standard deviation over seeds 42, 137, and 911.}}
\label{fig:fairness}
\end{figure}

{\color{black}
Table~\ref{tab:strong_baselines_comprehensive} compares
\textit{Fuzzy-FedAvg} with FedAvg, FedProx~\cite{li2020fedprox},
SCAFFOLD~\cite{karimireddy2020scaffold},
Power-of-Choice~\cite{cho2022client}, and Oort~\cite{lai2021oort}
using MFCC features, in terms of UAR, cumulative training time,
participation fairness (PF), accuracy fairness (AF), accuracy
disparity (AD), and worst-case accuracy (WCA). On IEMOCAP, \textit{Fuzzy-FedAvg} achieves a UAR of
$0.437 \pm 0.060$, exceeding FedAvg, FedProx, and Oort, whereas
SCAFFOLD ($0.501$) and Power-of-Choice ($0.503$) achieve higher UAR.
It reduces cumulative training time relative to FedAvg by
approximately 24.5\%, with a markedly smaller standard deviation, and
its PF ($0.935$) is close to that of the uniform-sampling methods and
far above that of Power-of-Choice ($0.562$). On CREMA-D, it achieves a
UAR of $0.418 \pm 0.029$, exceeding all baselines except SCAFFOLD
($0.479$), and records the lowest cumulative training time of all six
methods ($615.89 \pm 30.93$~s, 21.3--50.2\% below the baselines) while
maintaining a PF of $0.940$. No single method therefore dominates:
SCAFFOLD gains accuracy through explicit drift correction and
Power-of-Choice through a concentrated selection pattern, whereas
\textit{Fuzzy-FedAvg} balances UAR, training time, and participation.
Because drift correction acts on local optimisation and
\textit{Fuzzy-FedAvg} acts on participation and aggregation weights,
the two approaches are complementary.

Client-level fairness does not improve uniformly
(Table~\ref{tab:strong_baselines_comprehensive}
and Fig.~\ref{fig:fairness}). On IEMOCAP, FedAvg achieves lower AD
and higher WCA with both feature types, and Power-of-Choice attains the
best AD and WCA among all methods. On CREMA-D, \textit{Fuzzy-FedAvg}
yields marginally lower AD than FedAvg with MFCC features and improves
both indicators with Mel features (AD from $0.620$ to $0.555$; WCA from
$0.176$ to $0.217$). This is consistent with the design of the
framework: the CSWs prioritise accuracy, diversity, and timeliness,
none of which directly targets the worst-performing client.
}

\subsection{Effect of Statistical Heterogeneity ($\alpha$)}
\label{subsec:alpha}
\begin{figure*}[!ht]
\centering
\includegraphics[width=0.8\textwidth]{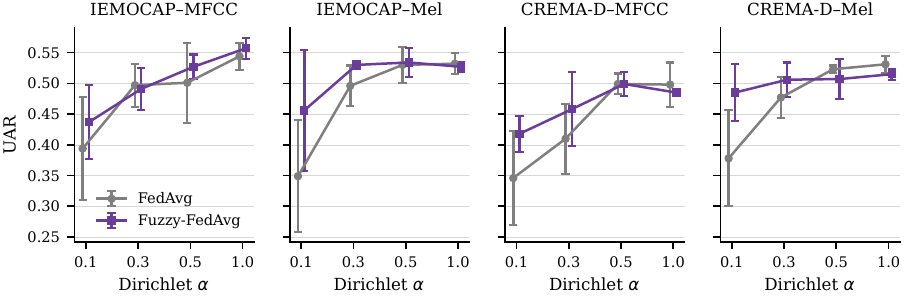}
\caption{\rv{UAR of FedAvg and \textit{Fuzzy-FedAvg} under varying statistical heterogeneity $\alpha$ ($K=30$, $N=8$, $G=80$). Markers show the mean and error bars the standard deviation over seeds 42, 137, and 911. \textit{Fuzzy-FedAvg} provides the clearest gains under severe non-IID conditions ($\alpha=0.1$), while the advantage becomes configuration-dependent as the data approach the IID regime.}}
\label{fig:alpha}
\end{figure*}

{\color{black}
Fig.~\ref{fig:alpha} reports UAR across
$\alpha \in \{0.1, 0.3, 0.5, 1.0\}$. At $\alpha=0.1$,
\textit{Fuzzy-FedAvg} achieves a higher mean UAR than FedAvg in all
four dataset--feature configurations, with gains of $0.043$ and
$0.107$ on IEMOCAP (MFCC and Mel) and $0.072$ and $0.107$ on CREMA-D
(MFCC and Mel). At $\alpha=0.3$, it remains higher in three of the
four configurations, and at $\alpha=0.5$ and $1.0$ the differences
become small and configuration-dependent. The benefit of adaptive
selection is therefore concentrated in the strongly non-IID regime,
where clients differ most in the information their updates carry; as
the data approach the IID regime, clients become more interchangeable
and the selection criteria offer less discrimination among them.
}

\begin{figure*}[!ht]
\centering
\includegraphics[width=0.8\textwidth]{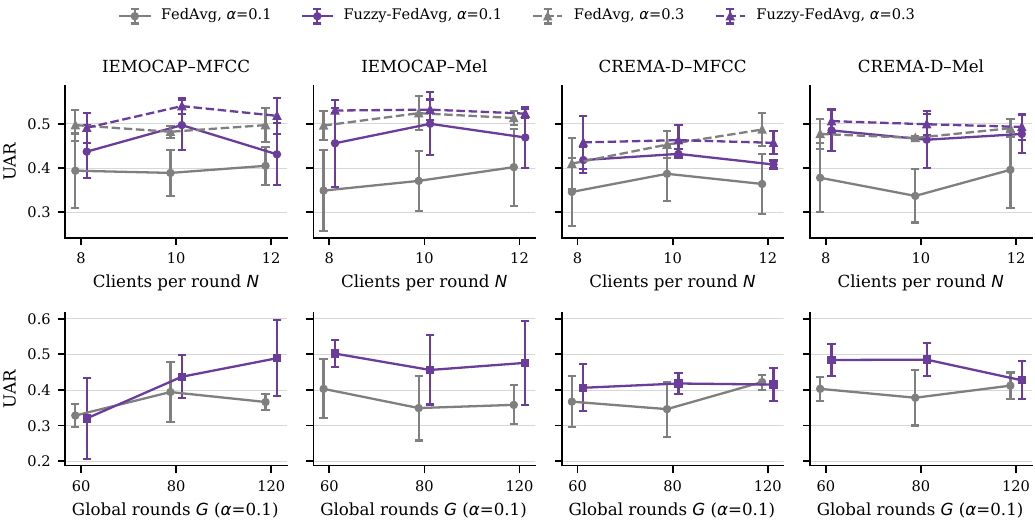}
\caption{\rv{UAR of FedAvg and \textit{Fuzzy-FedAvg} for (top row) varying numbers of participating clients per round $N$ at $\alpha \in \{0.1, 0.3\}$ with $G=80$, and (bottom row) varying numbers of global rounds $G$ at $\alpha=0.1$ with $N=8$; $K=30$ throughout. Markers show the mean and error bars the standard deviation over seeds 42, 137, and 911.}}
\label{fig:participation_horizon}
\end{figure*}

\begin{figure}[!ht]
\centering
\includegraphics[width=\columnwidth]{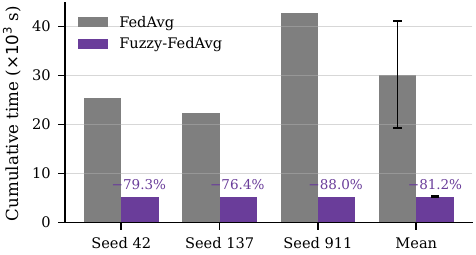}
\caption{\rv{Cumulative training time under FedScale-based heterogeneous client timing profiles (IEMOCAP, MFCC, $\alpha=0.1$, $K=30$, $N=8$, $G=80$) for each seed and on average (mean $\pm$ standard deviation). Labels give the reduction achieved by \textit{Fuzzy-FedAvg} relative to FedAvg.}}
\label{fig:timing}
\end{figure}
\subsection{Robustness to Participation Level and Training Horizon}
\label{subsec:clients}
\label{subsec:rounds}
{\color{black}

Fig.~\ref{fig:participation_horizon} (top row) reports UAR for
$N \in \{8,10,12\}$ at $\alpha \in \{0.1,0.3\}$.
\textit{Fuzzy-FedAvg} achieves a higher mean UAR than FedAvg in 22 of
the 24 configurations, including all twelve at $\alpha=0.1$ and all
eight at $N=10$, with the largest gains at $\alpha=0.1$ and $N=10$
(e.g., from $0.337$ to $0.464$ on CREMA-D with Mel features). Both
exceptions occur at $\alpha=0.3$. The bottom row
reports UAR for $G \in \{60,80,120\}$ at $\alpha=0.1$, where
\textit{Fuzzy-FedAvg} is higher in 10 of the 12 configurations. UAR
does not increase monotonically with either $N$ or $G$, and several
configurations show considerable run-to-run variability. This suggests
that the benefit depends on the interplay between the participation
budget, the training horizon, and the pool of eligible clients
remaining after cooldown and straggler exclusion, rather than growing
steadily with either parameter.
}

\begin{figure*}[!ht]
\centering
\includegraphics[width=0.9\textwidth]{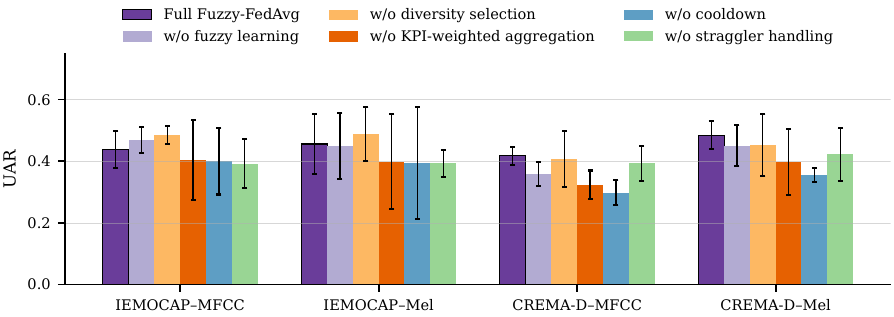}
\caption{\rv{Ablation study of the main components of \textit{Fuzzy-FedAvg} under severe statistical heterogeneity ($\alpha=0.1$, $K=30$, $N=8$, $G=80$). Bars show the mean UAR and error bars the standard deviation over seeds 42, 137, and 911.}}
\label{fig:ablation}
\end{figure*}

\subsection{System Efficiency}
\label{subsec:time}
{\color{black}
To validate the efficiency gains under more realistic system
conditions, a separate FedScale-based timing evaluation is conducted on
IEMOCAP with MFCC features in the reference configuration, using
heterogeneous client execution profiles that combine device-dependent
computation with throughput and network-delay variability. As shown in
Fig.~\ref{fig:timing}, \textit{Fuzzy-FedAvg} reduces
cumulative training time in all three runs, on average from
$30133.88 \pm 10938.92$~s to $5216.30 \pm 96.90$~s, a reduction of
$81.23 \pm 6.06\%$, with much more stable execution time across seeds.
These gains are not offset by the control mechanism: the fuzzy logic
and the PCAS selection and scheduling operations require approximately
$1.217$~ms and $6.331$~ms per round, respectively, amounting to about
$0.507$~s of PCAS overhead over $G=80$ rounds.

}

\subsection{Ablation Study}
\label{subsec:ablation}
{\color{black}
Fig.~\ref{fig:ablation} removes one component at a time.
Removing KPI-weighted aggregation, the cooldown mechanism, or
straggler handling reduces UAR in all four configurations, with the
largest reductions on CREMA-D (e.g., from $0.418$ to $0.298$ without
cooldown with MFCC features). These three components therefore
contribute consistently, confirming that coupling selection with
aggregation and preserving broad participation are central to the
framework. The effects of fuzzy learning and diversity-based selection
are dataset-dependent: removing either lowers UAR on CREMA-D, whereas
on IEMOCAP removing diversity-based selection increases UAR to
$0.485$ (MFCC) and $0.488$ (Mel), and removing fuzzy learning
increases it with MFCC features ($0.469$). A plausible explanation is
that the IEMOCAP training pool, drawn from three sessions and four
classes, offers limited variation for the diversity signal to exploit,
whereas CREMA-D, with six classes and many more speakers, provides
greater scope for diversity-aware selection.
}

\begin{table}[t!]
\centering
\caption{Sensitivity analysis of key Fuzzy-FedAvg design parameters on
IEMOCAP using MFCC features under severe statistical heterogeneity
($\alpha=0.1$), with $K=30$, $N=8$, and $G=80$. Each parameter is
varied while the remaining settings are held fixed at their nominal
values. Results correspond to seed 42.}
\label{tab:sensitivity_analysis}
\footnotesize
\setlength{\tabcolsep}{3pt}
\renewcommand{\arraystretch}{1.12}
{\color{black}
\begin{tabular}{p{2.1cm}lcc}
\hline
\textbf{Parameter} & \textbf{Setting} & \textbf{UAR $\uparrow$} & \textbf{Time (s) $\downarrow$} \\
\hline
\multirow{3}{=}{Conclusions}
& $\{0,2,4\}$ & $\mathbf{0.3920}$ & $387.55$ \\
& $\{0,3,6\}$ & $0.3399$ & $367.22$ \\
& $\{0,4,8\}$ & $0.3695$ & $\mathbf{366.79}$ \\
\hline
\multirow{3}{=}{Reward scale}
& $0.8$ & $0.3579$ & $357.46$ \\
& $1.0$ & $\mathbf{0.4369}$ & $\mathbf{356.96}$ \\
& $1.2$ & $0.4229$ & $384.53$ \\
\hline
\multirow{3}{=}{Cooldown}
& 1 round & $0.3700$ & $369.24$ \\
& 3 rounds & $\mathbf{0.4086}$ & $411.77$ \\
& 5 rounds & $0.3491$ & $\mathbf{283.90}$ \\
\hline
\multirow{3}{=}{PCAS Timing Threshold}
& Strict  & $0.4933$ & $\mathbf{413.88}$ \\
& Nominal & $0.4373$ & $417.05$ \\
& Relaxed & $\mathbf{0.5097}$ & $416.76$ \\
\hline
\multirow{3}{=}{Fuzzy membership}
& Lower   & $\mathbf{0.4536}$ & $\mathbf{412.85}$ \\
& Nominal & $0.4532$ & $414.53$ \\
& Upper   & $0.4191$ & $416.20$ \\
\hline
\multirow{2}{=}{Exploration strategy}
& Reward-dependent & $0.4734$ & $\mathbf{415.98}$ \\
& $\epsilon$-greedy & $\mathbf{0.5560}$ & $417.84$ \\
\hline
\end{tabular}
}
\end{table}

\subsection{Sensitivity Analysis}
\label{subsec:sensitivity}
{\color{black}
Table~\ref{tab:sensitivity_analysis} varies one design parameter at a
time on IEMOCAP with MFCC features, holding all other settings at their
nominal values and using a single controlled run with seed 42.

The conclusion set $\{0,2,4\}$ and the nominal reward scale of $1.0$
yield the highest UAR in their respective blocks, and the nominal
three-round cooldown gives the highest UAR among the cooldown lengths,
whereas a five-round cooldown gives the lowest cumulative time at a
cost in UAR. Both the relaxed and the strict timing thresholds exceed
the nominal setting, the lower and nominal fuzzy memberships perform
almost identically, and $\epsilon$-greedy exploration produces the
largest change, reaching $0.556$ compared with $0.473$ for
reward-dependent exploration. Except for the cooldown length, all
parameters affect UAR considerably more than training time. The
nominal configuration is therefore a reasonable operating point rather
than an optimum, and the exploration strategy and timing thresholds in
particular should be tuned for each deployment.
}

\FloatBarrier
\section{Discussion and Conclusion}
\label{sec:conclusion}
{\color{black}
This paper presented \textit{Fuzzy-FedAvg}, a server-aware
client-selection and aggregation framework for SER in heterogeneous
federated environments, in which a fuzzy Q-learning agent learns CSWs
from global-model accuracy, training time, and \divkpi{}, and a PCAS
node applies the same weights to client ranking and aggregation while
deferring near-on-time updates with a staleness discount. Taken together, the results in Section~\ref{sec:results} show that the value of adaptive client selection lies mainly in the strongly non-IID regime and in balancing competing objectives rather
than maximising any single metric. The framework improves UAR and
training efficiency over FedAvg while preserving broad participation,
but it does not systematically improve worst-client performance, and
two of its components, fuzzy adaptation and diversity-aware selection,
help consistently only on the dataset with greater speaker and class
variety.

Several limitations remain. The framework keeps raw speech on client
devices but does not incorporate formal privacy mechanisms such as
differential privacy or secure aggregation. The fairness metrics
describe participation and performance balance across simulated
clients, not demographic or group fairness. The evaluation relies on
simulated clients and controlled device profiles, and the baseline
comparison and sensitivity analysis were limited to MFCC features and
a single seed, respectively. Future work will address these points by
integrating privacy-preserving mechanisms, adding an explicit fairness
term to the selection objective, combining adaptive selection with
drift-correction methods such as SCAFFOLD, and extending the framework
to multilingual and cross-lingual SER.

The core findings of this study are as follows:
\begin{itemize}
\item Under severe non-IID conditions ($\alpha=0.1$),
\textit{Fuzzy-FedAvg} achieves higher UAR than FedAvg in all four
dataset--feature configurations, with gains of $0.043$--$0.107$.
\item It achieves higher UAR than FedAvg in 22 of 24
participation-level and 10 of 12 training-horizon configurations.
\item It outperforms FedAvg, FedProx, and Oort in UAR on both
datasets and achieves the lowest training time on CREMA-D, while
SCAFFOLD and Power-of-Choice remain competitive.
\item It reduces cumulative training time by $81.23 \pm 6.06\%$
relative to FedAvg in the FedScale-based evaluation, with negligible
control overhead.
\item KPI-weighted aggregation, cooldown, and straggler handling
contribute consistently, whereas fuzzy adaptation, diversity-aware
selection, and client-level fairness are dataset-dependent.
\end{itemize}
}

\section*{Declarations}
This work uses publicly available datasets which are cited in the relevant sections.\\

The authors declare no competing interests.\\

There was no funding obtained for this work.


\end{document}